\RequirePackage{lineno}
\documentclass[aps,prl,twocolumn,showpacs,superscriptaddress]{revtex4-1}  
\usepackage{graphicx}  
\usepackage{dcolumn}   
\usepackage{bm}        
\usepackage{amssymb}   

\usepackage{hyperref}
\usepackage{slashed}
\usepackage{multirow}

\newcommand{\met}{\ensuremath{\slashed{E}_T}}
\newcommand{\neutralino}{\ensuremath{\tilde{\chi}_1^0}}

\begin{document}

\hspace{5.2in} \mbox{FERMILAB-PUB-12-080-E}

\title{
Search for \texorpdfstring{$\boldsymbol{Z\gamma}$}{Zgamma} events with large missing transverse energy in \texorpdfstring{$\boldsymbol{p\bar{p}}$}{ppbar} collisions at \texorpdfstring{$\boldsymbol{\sqrt{s}}$}{sqrt(s)}~=~1.96~TeV
}
\affiliation{LAFEX, Centro Brasileiro de Pesquisas F\'{i}sicas, Rio de Janeiro, Brazil}
\affiliation{Universidade do Estado do Rio de Janeiro, Rio de Janeiro, Brazil}
\affiliation{Universidade Federal do ABC, Santo Andr\'e, Brazil}
\affiliation{University of Science and Technology of China, Hefei, People's Republic of China}
\affiliation{Universidad de los Andes, Bogot\'a, Colombia}
\affiliation{Charles University, Faculty of Mathematics and Physics, Center for Particle Physics, Prague, Czech Republic}
\affiliation{Czech Technical University in Prague, Prague, Czech Republic}
\affiliation{Center for Particle Physics, Institute of Physics, Academy of Sciences of the Czech Republic, Prague, Czech Republic}
\affiliation{Universidad San Francisco de Quito, Quito, Ecuador}
\affiliation{LPC, Universit\'e Blaise Pascal, CNRS/IN2P3, Clermont, France}
\affiliation{LPSC, Universit\'e Joseph Fourier Grenoble 1, CNRS/IN2P3, Institut National Polytechnique de Grenoble, Grenoble, France}
\affiliation{CPPM, Aix-Marseille Universit\'e, CNRS/IN2P3, Marseille, France}
\affiliation{LAL, Universit\'e Paris-Sud, CNRS/IN2P3, Orsay, France}
\affiliation{LPNHE, Universit\'es Paris VI and VII, CNRS/IN2P3, Paris, France}
\affiliation{CEA, Irfu, SPP, Saclay, France}
\affiliation{IPHC, Universit\'e de Strasbourg, CNRS/IN2P3, Strasbourg, France}
\affiliation{IPNL, Universit\'e Lyon 1, CNRS/IN2P3, Villeurbanne, France and Universit\'e de Lyon, Lyon, France}
\affiliation{III. Physikalisches Institut A, RWTH Aachen University, Aachen, Germany}
\affiliation{Physikalisches Institut, Universit\"at Freiburg, Freiburg, Germany}
\affiliation{II. Physikalisches Institut, Georg-August-Universit\"at G\"ottingen, G\"ottingen, Germany}
\affiliation{Institut f\"ur Physik, Universit\"at Mainz, Mainz, Germany}
\affiliation{Ludwig-Maximilians-Universit\"at M\"unchen, M\"unchen, Germany}
\affiliation{Fachbereich Physik, Bergische Universit\"at Wuppertal, Wuppertal, Germany}
\affiliation{Panjab University, Chandigarh, India}
\affiliation{Delhi University, Delhi, India}
\affiliation{Tata Institute of Fundamental Research, Mumbai, India}
\affiliation{University College Dublin, Dublin, Ireland}
\affiliation{Korea Detector Laboratory, Korea University, Seoul, Korea}
\affiliation{CINVESTAV, Mexico City, Mexico}
\affiliation{Nikhef, Science Park, Amsterdam, the Netherlands}
\affiliation{Radboud University Nijmegen, Nijmegen, the Netherlands}
\affiliation{Joint Institute for Nuclear Research, Dubna, Russia}
\affiliation{Institute for Theoretical and Experimental Physics, Moscow, Russia}
\affiliation{Moscow State University, Moscow, Russia}
\affiliation{Institute for High Energy Physics, Protvino, Russia}
\affiliation{Petersburg Nuclear Physics Institute, St. Petersburg, Russia}
\affiliation{Instituci\'{o} Catalana de Recerca i Estudis Avan\c{c}ats (ICREA) and Institut de F\'{i}sica d'Altes Energies (IFAE), Barcelona, Spain}
\affiliation{Uppsala University, Uppsala, Sweden}
\affiliation{Lancaster University, Lancaster LA1 4YB, United Kingdom}
\affiliation{Imperial College London, London SW7 2AZ, United Kingdom}
\affiliation{The University of Manchester, Manchester M13 9PL, United Kingdom}
\affiliation{University of Arizona, Tucson, Arizona 85721, USA}
\affiliation{University of California Riverside, Riverside, California 92521, USA}
\affiliation{Florida State University, Tallahassee, Florida 32306, USA}
\affiliation{Fermi National Accelerator Laboratory, Batavia, Illinois 60510, USA}
\affiliation{University of Illinois at Chicago, Chicago, Illinois 60607, USA}
\affiliation{Northern Illinois University, DeKalb, Illinois 60115, USA}
\affiliation{Northwestern University, Evanston, Illinois 60208, USA}
\affiliation{Indiana University, Bloomington, Indiana 47405, USA}
\affiliation{Purdue University Calumet, Hammond, Indiana 46323, USA}
\affiliation{University of Notre Dame, Notre Dame, Indiana 46556, USA}
\affiliation{Iowa State University, Ames, Iowa 50011, USA}
\affiliation{University of Kansas, Lawrence, Kansas 66045, USA}
\affiliation{Kansas State University, Manhattan, Kansas 66506, USA}
\affiliation{Louisiana Tech University, Ruston, Louisiana 71272, USA}
\affiliation{Boston University, Boston, Massachusetts 02215, USA}
\affiliation{Northeastern University, Boston, Massachusetts 02115, USA}
\affiliation{University of Michigan, Ann Arbor, Michigan 48109, USA}
\affiliation{Michigan State University, East Lansing, Michigan 48824, USA}
\affiliation{University of Mississippi, University, Mississippi 38677, USA}
\affiliation{University of Nebraska, Lincoln, Nebraska 68588, USA}
\affiliation{Rutgers University, Piscataway, New Jersey 08855, USA}
\affiliation{Princeton University, Princeton, New Jersey 08544, USA}
\affiliation{State University of New York, Buffalo, New York 14260, USA}
\affiliation{Columbia University, New York, New York 10027, USA}
\affiliation{University of Rochester, Rochester, New York 14627, USA}
\affiliation{State University of New York, Stony Brook, New York 11794, USA}
\affiliation{Brookhaven National Laboratory, Upton, New York 11973, USA}
\affiliation{Langston University, Langston, Oklahoma 73050, USA}
\affiliation{University of Oklahoma, Norman, Oklahoma 73019, USA}
\affiliation{Oklahoma State University, Stillwater, Oklahoma 74078, USA}
\affiliation{Brown University, Providence, Rhode Island 02912, USA}
\affiliation{University of Texas, Arlington, Texas 76019, USA}
\affiliation{Southern Methodist University, Dallas, Texas 75275, USA}
\affiliation{Rice University, Houston, Texas 77005, USA}
\affiliation{University of Virginia, Charlottesville, Virginia 22901, USA}
\affiliation{University of Washington, Seattle, Washington 98195, USA}
\author{V.M.~Abazov} \affiliation{Joint Institute for Nuclear Research, Dubna, Russia}
\author{B.~Abbott} \affiliation{University of Oklahoma, Norman, Oklahoma 73019, USA}
\author{B.S.~Acharya} \affiliation{Tata Institute of Fundamental Research, Mumbai, India}
\author{M.~Adams} \affiliation{University of Illinois at Chicago, Chicago, Illinois 60607, USA}
\author{T.~Adams} \affiliation{Florida State University, Tallahassee, Florida 32306, USA}
\author{G.D.~Alexeev} \affiliation{Joint Institute for Nuclear Research, Dubna, Russia}
\author{G.~Alkhazov} \affiliation{Petersburg Nuclear Physics Institute, St. Petersburg, Russia}
\author{A.~Alton$^{a}$} \affiliation{University of Michigan, Ann Arbor, Michigan 48109, USA}
\author{G.~Alverson} \affiliation{Northeastern University, Boston, Massachusetts 02115, USA}
\author{M.~Aoki} \affiliation{Fermi National Accelerator Laboratory, Batavia, Illinois 60510, USA}
% ...
%
%
\collaboration{The D0 Collaboration\footnote{with visitors from
$^{a}$Augustana College, Sioux Falls, SD, USA,
$^{b}$The University of Liverpool, Liverpool, UK,
$^{c}$UPIITA-IPN, Mexico City, Mexico,
$^{d}$DESY, Hamburg, Germany,
,
$^{e}$SLAC, Menlo Park, CA, USA,
$^{f}$University College London, London, UK,
$^{g}$Centro de Investigacion en Computacion - IPN, Mexico City, Mexico,
$^{h}$ECFM, Universidad Autonoma de Sinaloa, Culiac\'an, Mexico
and
$^{i}$Universidade Estadual Paulista, S\~ao Paulo, Brazil.
$^{\ddag}$Deceased.
}} \noaffiliation
\vskip 0.25cm

\date{March 23, 2012}

\begin{abstract}
We present the first search for supersymmetry (SUSY) in $Z\gamma$ final states with large missing transverse energy
using data corresponding to an integrated luminosity of $6.2$~fb$^{-1}$ collected with the D0 experiment
in $p\bar{p}$ collisions at $\sqrt{s} = 1.96$~TeV.
This signature is predicted in gauge-mediated SUSY-breaking models,
where the lightest neutralino $\tilde{\chi}_1^0$ is the next-to-lightest supersymmetric particle
and is produced in pairs, possibly through decay from heavier supersymmetric particles.
The $\tilde{\chi}_1^0$ can decay either to a $Z$ boson or a photon and an associated gravitino that escapes detection.
We exclude this model at the 95\% C.L.
for SUSY breaking scales of $\Lambda < 87$~TeV, corresponding to
neutralino masses of $M(\tilde{\chi}_1^0) < 151$~GeV.

\end{abstract}
\maketitle

Gauge-mediated SUSY breaking (GMSB)~\cite{gmsb} is a well-motivated 
model for physics beyond the standard model (SM). 
In GMSB models, SM gauge interactions serve as the messengers of SUSY breaking
and thereby the masses of the SUSY partners of SM particles are connected to the strength of their gauge interactions.
Assuming R-parity conservation, SUSY particles are produced in pairs,
each decaying to lighter states which always include the next-to-lightest supersymmetric particle (NLSP).
The final supersymmetric decay of the NLSP to SM particles and the nearly massless gravitino $\tilde{G}$
provide the typical signature used in GMSB searches.
A recently formulated model-independent framework 
for gauge mediation is discussed in Ref.~\cite{gmsb_model}.

The CDF, D0, ATLAS, CMS, and H1 Collaborations have all searched for GMSB neutralinos $\neutralino$
in the $\gamma\tilde{G}+\gamma\tilde{G}$ (and single $\gamma\tilde{G}$) final state,
assuming that the $\tilde{\chi}_1^0$ is the NLSP and decays promptly producing a photon~\cite{cdf_d0,atlas_cms,Aktas:2004cc}.
In this Letter, we present the first search for the $Z\tilde{G}+\gamma\tilde{G}$ final state.
The minimal GMSB model we consider is ``Model Line E'' of Ref.~\cite{GMSB_E}
which is characterized by six parameters:
the effective SUSY-breaking scale $\Lambda$ which is varied in the following,
the number of sets of messenger particles which is set to $n_5=2$,
the ratio of the Higgs vacuum expectation value which is chosen to be $\tan\beta=3$,
the mass of the messenger particles which is selected to be $M/\Lambda=3$,
the Higgs sector mixing parameter $\mu$ which is taken as $\mu=(3/4) M_1$
where $M_1$ is the hypercharge gaugino mass,
and the parameter $C_{\mathrm{grav}}$ which is linearly related to the gravitino mass
and is set to $C_{\mathrm{grav}}=1$~\cite{Baer:1998ve}.
In this model $\tilde{\chi}_1^0$ decays
with substantial branching fraction to $Z\tilde{G}$, as well as to $\gamma\tilde{G}$,
thereby providing a promising experimental signature
for the discovery of the $\tilde{\chi}_1^0$ NLSP in the $Z\tilde{G}+\gamma\tilde{G}$ final state.
The gravitinos escape detection,
leading to a $Z\gamma$ final state with large missing transverse energy, $\slashed{E}_T$.
We report a search for these events
in $p\bar{p}$ collisions recorded with the D0 detector~\cite{d0det} at the Fermilab Tevatron Collider.

The final state for this analysis contains a $Z$ boson decaying 
to $e^+e^-$ or $\mu^+\mu^-$, a photon of large transverse energy, and large $\met$.
The data have been collected using a set of inclusive electron or muon triggers, corresponding to 
an integrated luminosity of 6.2~$\pm$~0.4~fb$^{-1}$~\cite{d0lumi}.
The triggers have about 100\% (78\%) efficiency for signal in the $ee\gamma$ ($\mu\mu\gamma$) channel.

Electrons are required to have at least 90\% of their energy deposited 
in the electromagnetic (EM) calorimeter and
a distribution for EM shower consistent with that expected for an electron. They are further required to 
be isolated in both the calorimeter and the tracker.
A neural network (NN) multivariate discriminant~\cite{nn}, formed from the parameters of the EM shower and the track
associated with the electron candidate, as well as central preshower detector
information, is used to discriminate electrons from jets.
For electrons with $p_T=40$ GeV, the identification efficiency is $\approx 82\%$.

Muons are identified as track segments in the muon detector that match tracks found in the tracking system.
They must be synchronous with the beam crossing time to reject background 
from cosmic rays. Muons are also required to be isolated in both the calorimeter and the tracker.
The identification efficiency for muons with $p_T=40$ GeV is $\approx 79\%$.

Photons are identified in the central calorimeter (CC) and are required to be separated from leptons and 
jets by $\Delta R = \sqrt{(\Delta \eta)^2 + (\Delta \phi)^2} > 0.7$~\cite{geo}. 
Additional requirements are applied on the fraction of energy deposited in the EM calorimeter and on isolation in both 
the calorimeter and the tracker.
The shower width in the third layer (EM3) must be consistent with that of a photon. 
To suppress electrons misidentified as photons, the candidates must
not be spatially matched to a track or to energy depositions in the silicon microstrip or central fiber trackers
that lie along the trajectory of an electron~\cite{hitonroad}.
Further rejection of jets
is achieved with a NN discriminant similar to that used for electron selection. 
The average identification efficiency for photons with $p_T=40$~GeV is $\approx 75\%$.

The $\slashed{E}_T$ is the negative of the vectorial sum of
transverse components of energy depositions in the calorimeter,
corrected for identified photons, electrons, and muons. Jet energies are
calibrated using transverse energy balance in photon+jet
and dijet events~\cite{JES}, and these corrections are propagated to the calculation of $\slashed{E}_T$.

To select $Z\gamma+\slashed{E}_T$ events, we first require at least
two leptons that are consistent with originating from $Z \rightarrow e^+e^-$ or $\mu^+\mu^-$ decay. 
Each lepton must have $p_T>15$~GeV, with
one electron (muon) having $p_T>25\ (20)$~GeV.
The two leptons must have opposite charge and an invariant mass $ M_{\ell \ell}$
within the $Z$-mass windows of 78--104~GeV and 65--115~GeV  for the $ee$ and $\mu\mu$ channels, respectively.
A total number of 261,964 (306,541) $ee$ ($\mu\mu$) candidates satisfy these criteria.
We require at least one isolated photon with $p_T^\gamma >30$~GeV in the event.
To reduce background from photons radiated by the two leptons, we require a three-body invariant mass $M(\ell \ell \gamma)>120$~GeV, which results in a total number of 78 (91) $ee\gamma$ ($\mu\mu\gamma$) candidates.
The GMSB signal is expected in the region of large $\slashed{E}_T$.
We therefore require $\slashed{E}_T>30\ (40)$~GeV in the electron (muon) channel.
To remove events with spurious $\slashed{E}_T$ due to poorly reconstructed muons, 
we require that $\Delta\phi(\slashed{E}_T, \mu_1) < 2.85$
where $\mu_1$ is the highest-$p_T$ muon.
The $\met$ significance,
a likelihood discriminant based on the ratio of $\met$ and its uncertainty,
is required to be $> 5$.
These selections optimize sensitivity to signal.
No data is selected in the $ee\gamma+\slashed{E}_T$ final state,
and a single event is selected in the $\mu\mu\gamma + \slashed{E}_T$ final state.

The background to the $Z\gamma+\slashed{E}_T$ signal arises from
instrumental backgrounds caused by mismeasured $\met$, misidentified leptons or misidentified jets
in $Z\gamma$, $Z+$jets, $WW$, $WZ$, $ZZ$, $W+X$, $t\bar{t}$ and multijet processes.
The backgrounds are either estimated using control
samples in data or using Monte Carlo (MC) simulated events processed using a detailed {\sc  geant}-based simulation~\cite{geant} of the D0
detector response and overlaid with data from random beam crossings. The simulation is corrected for lepton identification
efficiencies and energy resolutions observed in data.

The SM $Z\gamma$ process is the dominant source of background. 
It is estimated using {\sc pythia}~\cite{pythia}. 
The photon $p_T$ spectrum from {\sc pythia} for initial state radiation (ISR) is corrected for QCD and electroweak next-to-leading order (NLO) effects
using the MC event generator of Ref.~\cite{baur_MC}.
The contribution from final state radiation in data is determined by fitting the $M(\ell\ell)$ distribution
of $Z\gamma$ MC events to data in the range $p_T^\gamma>10$~GeV and $\slashed{E}_T<30$~GeV
and is found to be
very small because of the requirements on $\Delta R(\ell,\gamma)$,
$p_T^\gamma$, $M(\ell\ell\gamma)$, and $\slashed{E}_T$.
We estimate the $Z\gamma$ contribution in the signal region to be
$0.23 \pm 0.05$~(stat) and $0.43 \pm 0.05$~(stat) events in the electron and muon channels, respectively.

Background from $Z+$jets events can enter the sample if a jet is misidentified as a photon and $\slashed{E}_T$ is large. 
Two data-driven methods are used to estimate this background. In the first method, we select 
an orthogonal sample of events with at least two electrons or two muons
and with a jet passing all
photon acceptance criteria except failing either the requirements on tracker isolation or on shower width in EM3.
The $Z+$jets background is then estimated by scaling this sample by an 
$\eta-$dependent factor $f$. 
This factor $f$ is the ratio of the probability for a jet to satisfy full photon-identification criteria to 
the probability to fail tracker isolation or shower width requirements.
It is measured using dijet data as a function of $\eta$ and $E_T$,
yielding typical values of 0.08 to 0.16 with uncertainties of 10\%. 
In the second method, the $Z$+jets background is estimated by fitting
the sum of the NN templates for photons and photon-like jets
to the observed photon NN distribution.
Templates of the NN distributions are obtained from simulations of photons and separately of jets, as the NN for data is found to be 
well modeled by MC~\cite{nn}.
The results from these two methods are consistent within their statistical uncertainties,
and the first method is used since it yields smaller uncertainties.
The resulting estimates of the $Z+$jets contribution in the signal region are
$0.09 \pm 0.08$~(stat) and $0.17 \pm 0.16$~(stat) in the electron and muon channels, respectively.

The multijet contribution to the background for $Z \rightarrow \ell \ell$ candidates is estimated
by fitting the $M_{\ell \ell}$ distribution using templates from jet-rich data and MC simulated $Z \rightarrow \ell \ell$ events.
Using weighted jet-rich data, the contribution in the signal region is found to be negligible.

The SM backgrounds from $WW$, $WZ$, $ZZ$, and $t\bar{t}$ production are estimated using MC simulations.
The $\met$ can be substantial in such events, but none of these backgrounds are sources of isolated, high-$p_T^\gamma$ photons.
The contribution from $t\bar{t}$ events is minimized by the requirements on $M_{\ell \ell}$.

The GMSB signal is modeled with the {\sc pythia}
leading-order (LO) MC event generator using supersymmetric particle spectra calculated in 
{\sc isajet}~\cite{isajet}.
The $\Lambda$ parameter is varied from 70~TeV to 95~TeV, in steps of 5~TeV,
and used to compute an MSSM particle mass spectrum and a set of branching ratios. 
The LO signal cross sections are scaled to match the NLO prediction from {\sc prospino}~\cite{prospino}.
The inclusive cross section for the pair production of $\neutralino$ from cascade decays
is 618~fb for $\Lambda=70$ TeV and decreases to 106~fb for $\Lambda=95$ TeV.
The fraction of $\neutralino \rightarrow Z\tilde{G}$ decays ({\cal B}$_Z$) increases with $\Lambda$,
reaching 50\% at $\Lambda\approx 85$~TeV.  
Cross~sections and branching fractions are given in Table~\ref{tab:xsecbr}.
At larger $\Lambda$ values, $Z\tilde{G}$ is the main decay mode for $\neutralino$.
For the full event selection, the overall product of acceptance and efficiency is
7.7 (5.1)\% at $\Lambda=70$~TeV and increases to 11.2 (8.6)\% for $\Lambda=95$~TeV in the electron (muon) channel.

\begin{table}
  \caption{
     Cross~sections $\sigma_\mathrm{p}$ for the production of pairs of lightest neutralinos $\neutralino$ via cascade decay,
     branching fractions of $\neutralino$ to $\gamma\tilde{G}$ ({\cal B}$_\gamma$) and to $Z\tilde{G}$ ({\cal B}$_Z$),
     and the lightest neutralino mass $M_{\neutralino}$ used in this analysis,
     which is parametrized by the breaking scale~$\Lambda$.
     The $\neutralino$ also decays to Higgs$+\tilde{G}$ and to nonresonant $\ell^+\ell^-\tilde{G}$,
     which dominate the remaining decays for large and small $\Lambda$, respectively.
     Also given are the observed (expected) 95\% C.L. upper limits on the
     production cross~section.
  }
  \label{tab:xsecbr}
  \begin{ruledtabular}
  \begin{tabular}{cccccc}
    $\Lambda$ &
    $\sigma_\mathrm{p}$ &
    {\cal B}$_\gamma$ &
    {\cal B}$_Z$ &
    $M_{\neutralino}$ &
    obs. (exp.) limit \\
    {[}TeV] &
    [fb] &
    &
    &
    [GeV] &
    on $\sigma_\mathrm{p}$ [fb] \\
    \noalign{\smallskip}\hline\noalign{\smallskip}\noalign{\smallskip}
     70 & 618 & 0.892 & 0.086 & 111 & $< 234$ (223) \\
     75 & 419 & 0.715 & 0.253 & 123 & $< 172$ (150) \\
     80 & 290 & 0.545 & 0.408 & 135 & $< 167$ (140) \\
     85 & 205 & 0.420 & 0.519 & 147 & $< 163$ (137) \\
     90 & 146 & 0.335 & 0.592 & 159 & $< 186$ (155) \\
     95 & 106 & 0.277 & 0.642 & 169 & $< 205$ (159) \\
  \end{tabular}
  \end{ruledtabular}
\end{table}

\newcolumntype{d}{D{.}{.}{12}}

\begin{table*}
  \caption{
     Number of observed and expected events for the restrictive criteria defining the signal region and for less stringent requirements that are followed by a selection on BDT output
     defining an alternative signal region.
     The first uncertainty is statistical and the second is systematic.
     The contributions from $Z+$jets for the BDT-analyses are found to be negligible.
  }
  \label{tab:signal_bkg}
  \begin{ruledtabular}
  \begin{tabular}{ldddd}
            & \multicolumn{2}{c}{$ee\gamma+\met$}                                   & \multicolumn{2}{c}{$\mu\mu\gamma+\met$}  \\
            & \multicolumn{1}{c}{Signal region} & \multicolumn{1}{c}{BDT $>$ 0.8}   & \multicolumn{1}{c}{Signal region} & \multicolumn{1}{c}{BDT $>$ 0.8} \\
    \noalign{\smallskip}\hline\noalign{\smallskip}\noalign{\smallskip}
    Signal ($\Lambda=80$~TeV)  &  3.28 \,\pm\, 0.09 \,\pm\, 0.24  &  3.95 \,\pm\, 0.10 \,\pm\, 0.50  &  2.42 \,\pm\, 0.08 \,\pm\, 0.31  &  2.69 \,\pm\, 0.08 \,\pm\, 0.33  \\
    Signal ($\Lambda=90$~TeV)  &  1.48 \,\pm\, 0.03 \,\pm\, 0.11  &  1.73 \,\pm\, 0.05 \,\pm\, 0.21  &  1.06 \,\pm\, 0.03 \,\pm\, 0.14  &  1.22 \,\pm\, 0.04 \,\pm\, 0.15  \\
    \hline
    $Z\gamma$                  &  0.23 \,\pm\, 0.05 \,\pm\, 0.02  &  0.23 \,\pm\, 0.11 \,\pm\, 0.02  &  0.43 \,\pm\, 0.05 \,\pm\, 0.40  &  0.10 \,\pm\, 0.03 \,\pm\, 0.20  \\
    $Z+$jet                    &  0.09 \,\pm\, 0.08 \,\pm\, 0.01  &  \multicolumn{1}{c}{-}   &  0.17 \,\pm\, 0.16 \,\pm\, 0.02  &  \multicolumn{1}{c}{-}   \\
    $WW+WZ+ZZ$                 &  0.13 \,\pm\, 0.05 \,\pm\, 0.01  &  0.06 \,\pm\, 0.04 \,\pm\, 0.01  &  0.08 \,\pm\, 0.03 \,\pm\, 0.01  &  0.16 \,\pm\, 0.19 \,\pm\, 0.02  \\
    $t\bar{t}$                 &  0.05 \,\pm\, 0.01 \,\pm\, 0.01  &  0.14 \,\pm\, 0.03 \,\pm\, 0.02  &  0.04 \,\pm\, 0.01 \,\pm\, 0.01  &  0.05 \,\pm\, 0.02 \,\pm\, 0.01  \\
    All backgrounds            &  0.50 \,\pm\, 0.11 \,\pm\, 0.03  &  0.43 \,\pm\, 0.12 \,\pm\, 0.03  &  0.71 \,\pm\, 0.17 \,\pm\, 0.40  &  0.31 \,\pm\, 0.10 \,\pm\, 0.20  \\
    \hline
    Data                       &  \multicolumn{1}{c}{0}   &  \multicolumn{1}{c}{0}   &  \multicolumn{1}{c}{1}   &  \multicolumn{1}{c}{1}   \\
  \end{tabular}
  \end{ruledtabular}
\end{table*}

\begin{figure*}
\includegraphics [width=0.44\textwidth] {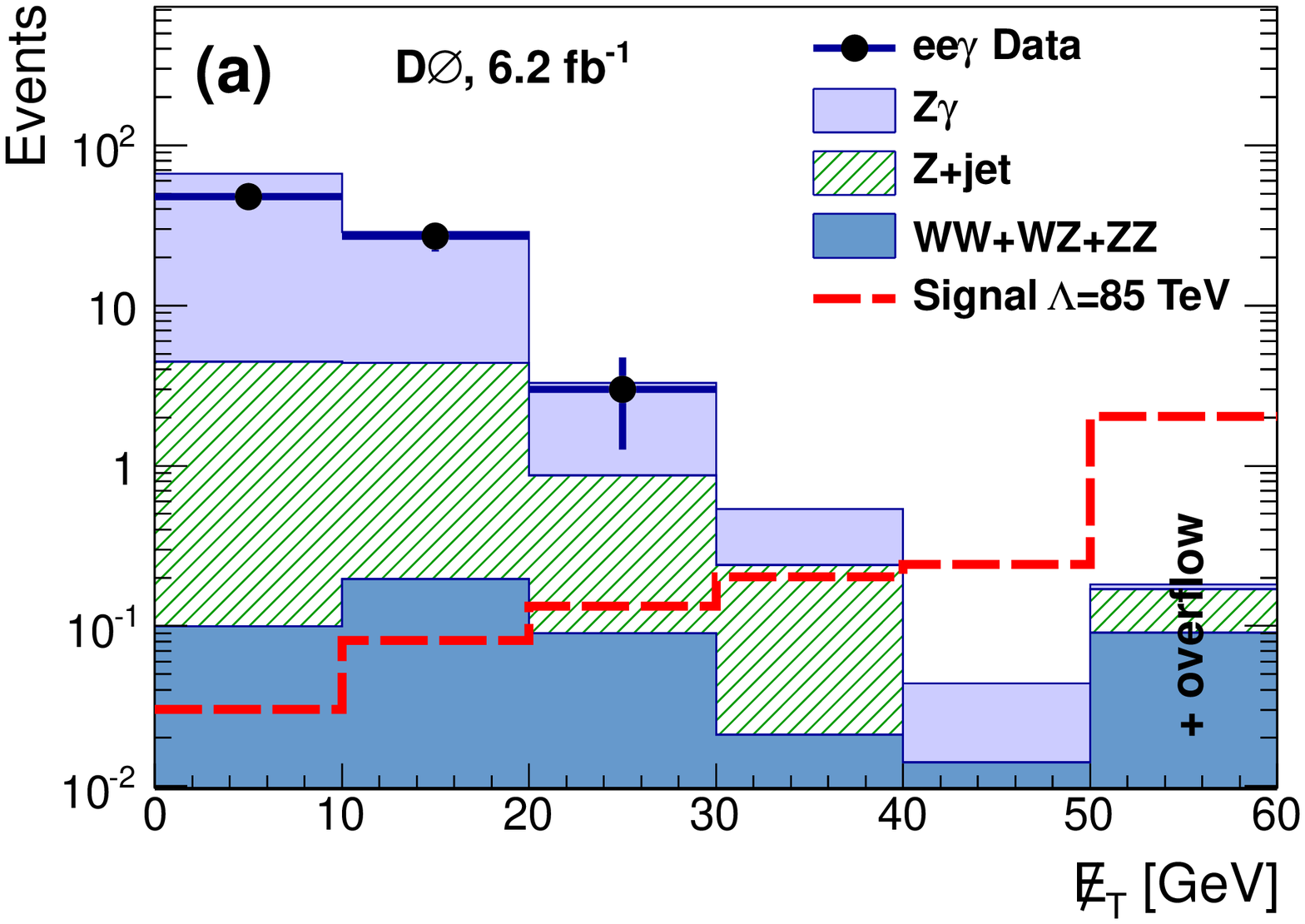}
\hspace*{1.0cm}
\includegraphics [width=0.44\textwidth] {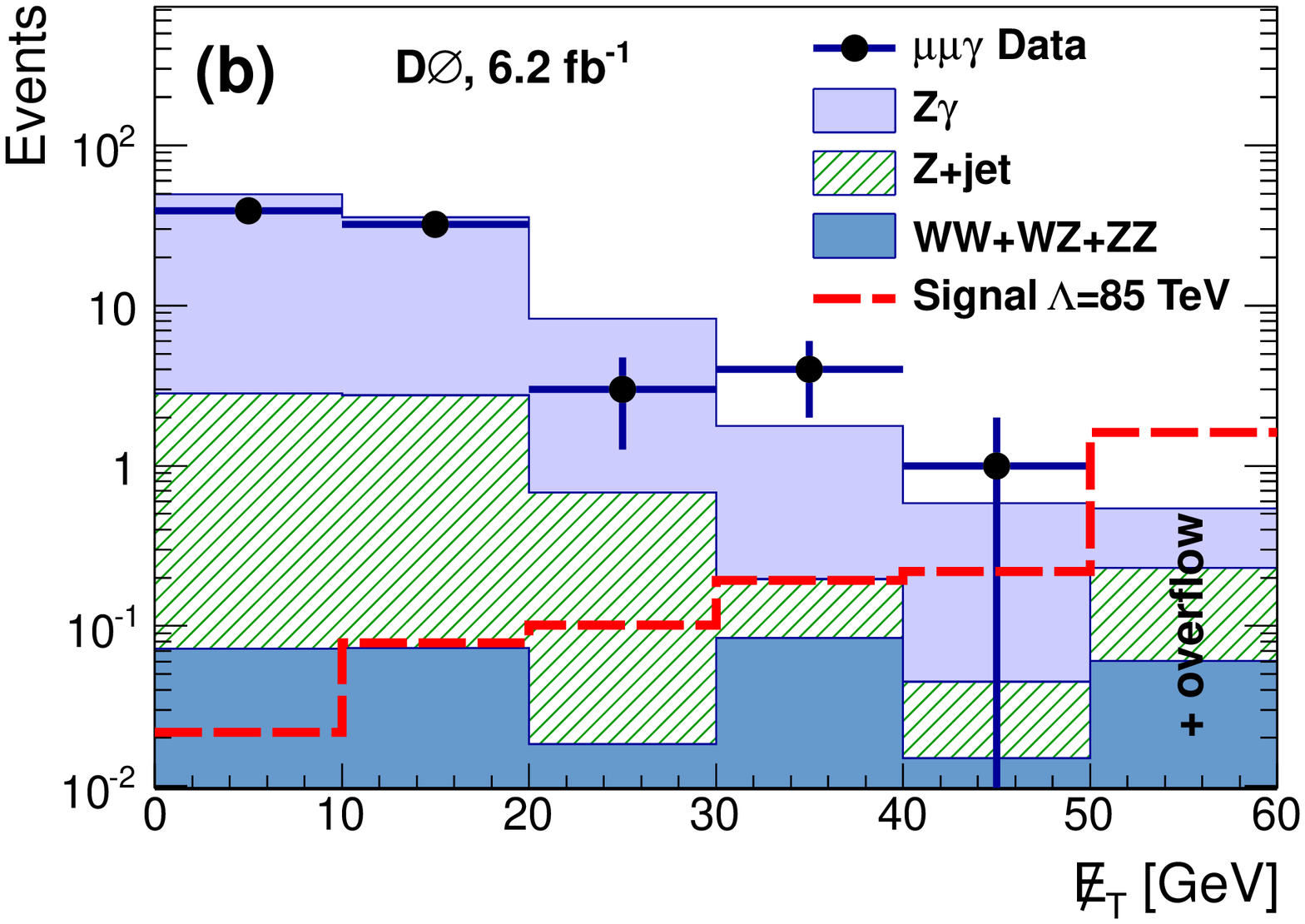}
\caption{
    \label{fig:met_comparison}
    Distribution of $\slashed{E}_T$ for $Z\gamma$ events in the (a) $ee\gamma$ channel and (b) $\mu\mu\gamma$ channel
    before requiring $\met$ significance~$> 5$.
}
\end{figure*}

The expected signal yield for $\Lambda=80$ and $90$~TeV and the estimated SM backgrounds are summarized in Table~\ref{tab:signal_bkg}. 
The total background is expected to be 0.5 $\pm$ 0.1
and 0.7 $\pm$ 0.4 events in the $ee\gamma+\slashed{E}_T$ and $\mu\mu\gamma+\slashed{E}_T$ channels, respectively.
The number of observed events is consistent with these expectations.
The comparison between data and SM MC predictions for the $\slashed{E}_T$ distributions after selecting $Z\gamma$ events
is given in Fig.~\ref{fig:met_comparison} along with the signal expectation.
Good agreement between data and SM background is observed for both $ee\gamma$ and $\mu\mu\gamma$ channels.

The systematic uncertainties that affect the signal and
SM backgrounds include theoretical and experimental sources.
The uncertainties on the theoretical cross~section for diboson and $t\bar{t}$ processes
are 6\% and 10\%, respectively.
The uncertainty on the measured luminosity is 6.1\%~\cite{d0lumi} and is applied
to the SM background estimations based on MC simulation.
The uncertainty on electron identification efficiency is 1\% in the CC region
and increases to 4\% in the end-cap calorimeter.
The systematic uncertainties on muon identification include 1.0\% for reconstruction, 
1.1\% for tracking efficiency, and 0.5\% for isolation.
The photon identification uncertainty is 2.7\%.
The uncertainties from the jet energy scale are estimated to be
1\% for signal and $4$\% for the backgrounds.
The uncertainty on the momentum resolution for muons is reflected in an uncertainty of $\approx 100\%$ in the signal region
$\met>40$~GeV on the estimate of the background from $Z(\mu\mu)+\gamma$.

\begin{figure}
\includegraphics [width=0.44\textwidth] {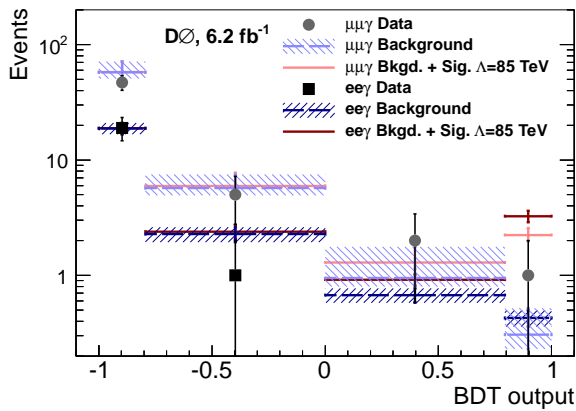}
\caption{
    \label{fig:bdt_comparison}
    Distribution of BDT output in the $ee\gamma$ channel and $\mu\mu\gamma$ channel
    for background only, background with a $\Lambda=85$~TeV signal added, and for data.
    The total background uncertainties are indicated as shaded bands.
}
\end{figure}

To improve the sensitivity for $\neutralino$ detection at the cost of a stronger dependence on the specifics of the GMSB model,
we also use a BDT multivariate technique to discriminate between SM background and signal~\cite{BDT}.
The output is a discriminant that is shifted toward $+1$ for signal,
and strongly peaked near $-1$ for background events.

The BDT is trained on a randomly selected collection of signal and background MC events,
$Z+$jet background candidates from data,
and a signal assuming $\Lambda = 90$~TeV.
The training samples require a leading lepton of $p_T> 25\ (20)$~GeV, a second lepton of $p_T > 15$~GeV, $p_T^\gamma>20$ GeV, $M(\ell \ell \gamma)>120$ GeV, $\slashed{E}_T>15$~GeV and 
$M(\ell \ell)>70\ (65)$~GeV in the electron (muon) channel.
A set of 14 sensitive variables, well modeled by the simulation, is used to form the BDT discriminant.
The variables include transverse momenta of the two leptons, photon, dilepton system, and 
dilepton+photon system, as well as $\slashed{E}_T$ and $M(\ell \ell \gamma)$.
The expected signal and background yields are estimated from events independent of the set used for training.
The data is found consistent with the SM background prediction
as seen in Fig.~\ref{fig:bdt_comparison} and Table~\ref{tab:signal_bkg} (for BDT $> 0.8$),
and no evidence is observed for a GMSB neutralino NLSP.

\begin{figure}
\includegraphics [width=0.44\textwidth] {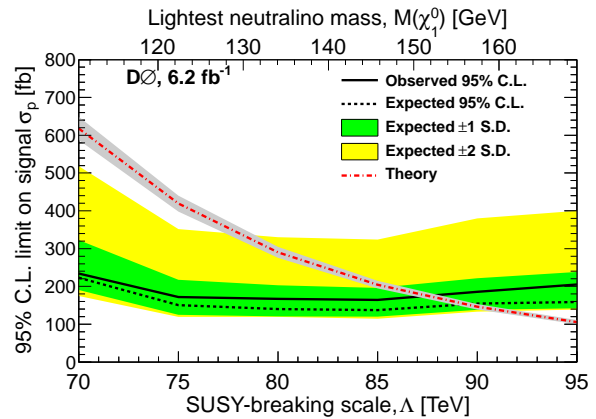}
\caption{
    Limit on the cross~section for
    $Z\gamma+\slashed{E}_T$ production
    as a function of $\Lambda$ (lower horizontal axis)
    and $M(\tilde{\chi}_1^0)$ (upper horizontal axis) at 95\%~C.L.
    combined for the $ee\gamma$ and $\mu\mu\gamma$ channels.
    The NLO cross~section from theory is overlaid.
    \label{fig:LimitFromCutsChannels}
}
\end{figure}

Limits on the production cross section of $\neutralino\neutralino$ using the benchmark model 
are derived using a Poisson log-likelihood ratio as test statistic, combining results from the electron and muon channels.
Pseudo-experiments are generated according to the background-only and signal+background hypotheses,
and systematic uncertainties are accounted for by integrating over uncertainties parametrized as Gaussian.
The limits on cross~sections are evaluated using the modified frequentist approach~\cite{CLs_method}. 
Data and background estimates are studied in four bins of BDT output,
and values from the most signal-like bin (BDT $> 0.8$) are shown in Table~\ref{tab:signal_bkg}.
The 95\% C.L. upper limit on the cross~section 
using the BDT discriminant is shown in Fig.~\ref{fig:LimitFromCutsChannels}, together with the expected limit 
and the 1 and 2 standard deviation (SD) uncertainty bands.
The 95\% C.L. limits on $\sigma_p$ are also given in Table~\ref{tab:xsecbr}.
Scales of $\Lambda < 87~\text{TeV}$ are excluded at 95\% C.L. which corresponds to $M(\neutralino) < 151~\text{GeV}$.

In summary, we present the first search for a SUSY signature
in events containing $Z\gamma+\slashed{E}_T$ final states using $6.2$~fb$^{-1}$ 
of integrated luminosity collected by the D0 experiment in $p\bar{p}$ collisions at $\sqrt{s} = 1.96$~TeV.
The signature corresponds to a GMSB model
where pairs of neutralino NLSPs are either produced promptly or from decays of other supersymmetric particles 
in $p\bar{p}$ collisions
and then decay to either $Z\tilde{G}$ or $\gamma\tilde{G}$.
In the expected signal region we observe no event in the $ee\gamma+\slashed{E}_T$ and one event in the $\mu\mu\gamma+\slashed{E}_T$
channels, where the SM background is expected to be $1.21 \pm 0.45$ combined.
Employing a multivariate selection process and combining the results from both channels,
the specific neutralino NLSP model is excluded at the 95\% C.L.
for $\Lambda < 87$~TeV, corresponding to
neutralino masses of $M(\neutralino) < 151$~GeV.

%
We thank the staffs at Fermilab and collaborating institutions,
and acknowledge support from the
DOE and NSF (USA);
CEA and CNRS/IN2P3 (France);
MON, Rosatom and RFBR (Russia);
CNPq, FAPERJ, FAPESP and FUNDUNESP (Brazil);
DAE and DST (India);
Colciencias (Colombia);
CONACyT (Mexico);
NRF (Korea);
FOM (The Netherlands);
STFC and the Royal Society (United Kingdom);
MSMT and GACR (Czech Republic);
BMBF and DFG (Germany);
SFI (Ireland);
The Swedish Research Council (Sweden);
and
CAS and CNSF (China).


\begin{thebibliography}{99}
\bibitem{geo} The polar angle $\theta$, the azimuthal angle $\phi$, and transverse quantities such as transverse momentum $p_T$ are defined 
with respect to the $z$-axis, which is along the proton beam direction, and the center of the D0 detector.
Pseudorapidity is defined as $\eta=-\ln[\tan(\theta/2)]$.
\end{thebibliography}
\end{document}